# Magnetic Phase Diagrams of Antiferromagnet $DyB_{12}$ with Jahn-Teller Lattice Instability and Electron Phase Separation

A.N. Azarevich[1], A.V. Bogach[1], K.M. Krasikov[1], V.V. Voronov[1], S.Yu. Gavrilkin[2], A.Yu. Tsvetkov[2], S. Gabani[3], K. Flachbart[3], N.E. Sluchanko[1,*]

*[1]Prokhorov General Physics Institute, Russian Academy of Sciences, Vavilov str. 38, Moscow 119991, Russia*

*[2]Lebedev Physical Institute, Russian Academy of Sciences, Moscow, Leninsky Prospect 53, 119991 Russia*
*[3]Institute of Experimental Physics of the Slovak Academy of Sciences, Watsonova 47, SK-04001 Košice, Slovakia*

[*] *e-mail: nes@lt.gpi.ru*

**Abstract.** The origin of charge transport and magnetization anisotropy was studied in $DyB_{12}$, an antiferromagnetic (AF) metal with Néel temperature $T_N \approx 16.3$ K that exhibits both cooperative Jahn-Teller distortions of the *fcc* crystal structure and nanoscale electronic instabilities (dynamic charge stripes). Based on the results obtained the magnetic field ($H$) vs temperature ($T$) phase diagrams have been constructed. Moreover, from angle ($\varphi$) dependent magnetoresistance and magnetization measurements the *butterfly-type* patterns of the $H$-$\varphi$ magnetic phase diagram in the (110) plane were created, which include a number of different magnetic phases separated from each other by radial and circular boundaries. Several positive and negative contributions to magnetoresistance were separated and analyzed, providing arguments in favor of the important role of the spin density wave 5*d*- component in the magnetic structure of AF state. We argue that charge fluctuations in stripes are responsible for the suppression of the Ruderman-Kittel-Kasuya-Yoshida (RKKY) indirect exchange between the nearest neighbored $Dy^{3+}$ ions located along the same ⟨110⟩ directions, as these dynamic charge stripes produce the magnetic phase diversity and the butterfly-type anisotropy in $DyB_{12}$.

***Keywords:*** *butterfly-type magnetic anisotropy, antiferromagnetic phase diagrams, RKKY exchange, dynamic charge stripes*

## I. Introduction

Rare earth (RE) dodecaborides $RB_{12}$ with a cage-glass structure [1] attract considerable attention due to unique combination of their characteristics, including high chemical stability, high melting temperature, microhardness, etc. In these antiferromagnetic (AF) metals the Néel temperature decreases monotonously from $TbB_{12}$ ($T_N \approx 22$ K) to $TmB_{12}$ ($T_N \approx 3.2$ K) in the $RB_{12}$ series while maintaining similar conduction band, composed from 5*d* (R) and 2*p* (B) atomic orbitals, and changing only the filling of the 4*f* shell of RE ions in the range $8 \leq n_{4f} \leq 14$ [2-4]. In these RE antiferromagnets the Ruderman–Kittel–Kasuya-Yosida (RKKY) interaction (the indirect exchange via conduction electrons) couples the magnetic moments of 4*f* orbitals. The long range and oscillatory character of this coupling in the presence of crystal electric field anisotropy, magnetoelastic coupling, dipole-dipole, or two-ion quadrupolar interactions in combination with many-body effects leads to competing interactions between RE ions and resulting in complicated magnetic structures and diagrams. Hence, these AF metals with unfilled 4*f* – shell and incommensurate helical or amplitude-modulated magnetic structures, having the simple face centered cubic (*fcc*) crystal structure with single type magnetic ions located in high-symmetry positions (see Fig.1a), look at first glance like very suitable systems for testing the theories and models developed for the magnetic properties of RE intermetallics. In this paradigm, the stable $B_{12}$ nanoclusters have been considered as basic structural elements of the *fcc* lattice of dodecaborides in band structure calculations. A new approach was developed recently in the studies of $RB_{12}$ [5-6] consisting on the difference Fourier synthesis of the residual electron density (ED) as well as on the construction of the ED distribution using the maximum entropy method (MEM). It was found [5-6] that in $RB_{12}$ various types of ED singularities exist and become stronger with the temperature lowering, forming filamentary structures of conduction channels – unbroken *dynamic charge stripes* almost along one of the <110> axes in the *fcc* lattice of RE dodecaborides (Fig.1a). The reason of the nanoscale electron phase separation is the development of *a cooperative dynamic Jahn-Teller (JT) instability of $B_{12}$ clusters*, leading to the emergence of infra-red active collective excitations, which involve JT-modes and conduction electrons. The static and dynamic JT effects produce periodic changes of the hybridization of 5*d*-2*p* states causing modulation of electron density in the conduction band and creating a large fraction (~70%) of non-equilibrium (hot) conduction electrons with very strong charge carrier scattering [7-8].

The fluctuating charges in stripes along <110> are also among the main factors responsible for the suppression of the RKKY interaction between the nearest neighbored magnetic RE-ions located at a distance of 5.3 Å from each other in the same <110> directions, modifying strongly the magnetic phase diagrams of $REB_{12}$ antiferromagnets [3]. It was shown in studies of $Ho_xLu_{1-x}B_{12}$ [9-11], $ErB_{12}$ [12] and $Tm_{1-x}Yb_xB_{12}$ [13-14], that the suppression of the RKKY interaction by nanoscale filamentary ED (stripes) should be considered as the main cause of the electron instability, providing both a symmetry lowering and the formation of Maltese-cross-like magnetic *H*-*φ* phase diagrams of $Ho_xLu_{1-x}B_{12}$ [9-11] and $Tm_{1-x}Yb_xB_{12}$ [13-14], and butterfly-type ones in $ErB_{12}$ [12]. A number of different magnetically ordered AF-phases in the RE dodecaborides with Néel temperatures $T_N < 8$ K were derived from precise angle-resolved magnetoresistance (ARMR) and magnetization measurements [3, 9-14]. In this study we have investigated dysprosium dodecaboride ($DyB_{12}$) with a much higher $T_N \approx 16.3$ K [2-4], which contains Kramers-type RE-ions similar to the situation in $ErB_{12}$, where the unusual antiferromagnetic ground state is accompanied by electronic phase separation [3-4]. The obtained results allowed us to construct for the first time both the principal *H*-*T* and the butterfly-type *H*-*φ* magnetic phase diagrams in the AF state of $DyB_{12}$, providing arguments in favor of the role of symmetry of the 4*f*-multiplet ground state in the formation of the Maltese-cross-like or butterfly-type anisotropy.

## II. Experimental details

Detailed studies of magnetization and magnetoresistance (MR) in combination with heat capacity measurements were performed on high-quality single-domain single crystals of $DyB_{12}$ that were grown by induction zone melting in an inert gas atmosphere [15]. The magnetoresistance measurements were carried out on an original setup, which allows rotating step-by-step the sample around the current axis $\mathbf{I}||[110]$ in external magnetic field up to 80 kOe in the transverse configuration ($\mathbf{I}\perp\mathbf{H}$), when $\mathbf{H}$ orientation varies in the (110) plane, passing principal directions [001], [110] and [111] in *fcc* lattice. The location of phase boundaries in field-temperature (*H*-*T*) and field-angle (*H*-*φ*) magnetic diagrams ($\angle\varphi = \mathbf{H}$^$\mathbf{n}$ and $\mathbf{n}$ || [001] is the normal vector to the lateral surface of the sample, see Fig. 1b) was confirmed by low temperature (1.8–20 K) measurements of magnetization (*M*) and heat capacity (*C*) in magnetic field up to 50 and 90 kOe, using correspondingly commercial Quantum Design installations MPMS-5 and PPMS-9. The MPMS-5 was used also to measure the magnetization angular dependences in the plane $\mathbf{H}$ || (110). The sample mount setup provides high precision (~1%) absolute magnetization values, which consider the demagnetizing factor, finite sample size and radial displacement inside the SQUID-magnetometer pickup coils.

## III. Results and discussion

**III.1. AF transition at *H* ≈ 0.** The AF-P (P-paramagnetic state) phase transition at $T_N \approx 16.3$ K [2-4] in $DyB_{12}$ are discerned clearly on the temperature dependences in small field ($H_0$=100 Oe) magnetization, zero field resistivity (Fig. 1c) and specific heat (Fig. 1d). It is worth to note that the low-field magnetization in the AF state is about isotropic (see curves for ***H***||[001], ***H***||[110], ***H***||[111] and ***H***||[112] in Fig. 1c), indicating a very strong spin disorder on a short-range scale (≤ 5 lattice constants), which is usually typical for a polycrystalline sample of an ordered magnet, and is similar to the results deduced previously for single crystals of $ErB_{12}$ [12,16]. Additionally, the hump observed in the resistivity curve just below $T_N$ is attributed to the formation of a spin density wave (SDW), being a 5*d*-component of the magnetic structure in the AF ordered state of $DyB_{12}$. In this case, the sharp increase of resistivity $\rho(T)$ below $T_N$ is caused by involvement of conduction electrons in the SDW arranged in the AF phase (see e.g. the itinerant anti-ferromagnetism in chromium [17]), and it was also discussed previously for $Ho_xLu_{1-x}B_{12}$ [9-11, 18]. According to [4], the observed behavior of the heat capacity in the critical region $T \leq T_N$ is typical for amplitude-modulated magnetic systems with a strong rise of $C(T)$ at the Néel temperature and a maximum below $T_N$ (Fig. 1d). Neutron diffraction experiments have revealed that the ground state in magnetic RE dodecaborides is frustrated, exhibiting multi-*q* incommensurate AF structures, and characterized by propagation vectors $\boldsymbol{q}_{AF} = (1/2 \pm \delta, 1/2 \pm \delta, 1/2 \pm \delta)$ for $Ho^{11}B_{12}$ and $Tm^{11}B_{12}$, and $\boldsymbol{q}_{AF} = (3/2 \pm \delta, 1/2 \pm \delta, 1/2 \pm \delta)$ for $Er^{11}B_{12}$ with $\delta = 0.035$ (see, for example, [19])]. Note that those types of AF arrangement in real space are similar to SDW observed in pure chromium and its alloys [20,21].

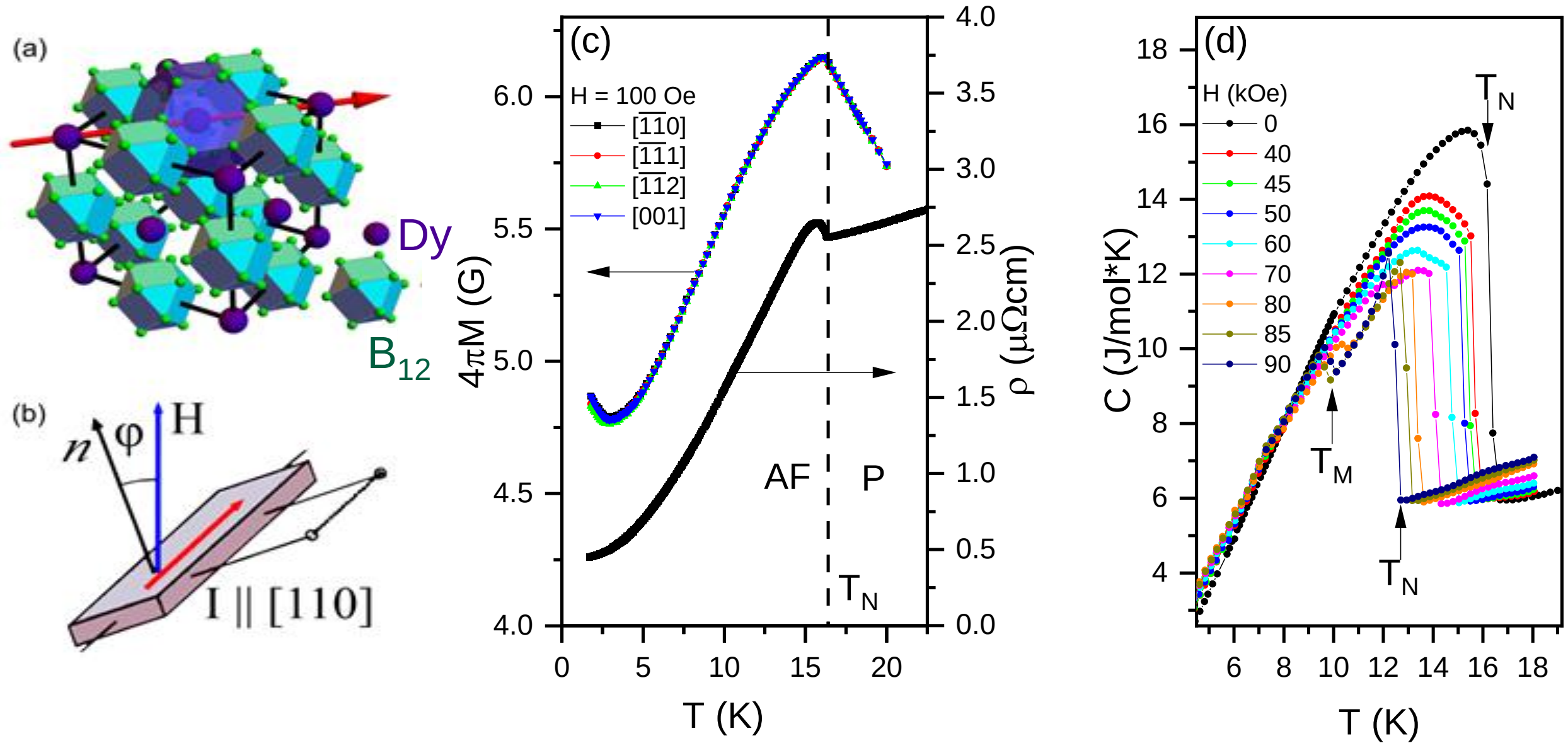

**Fig. 1. (a)** Crystal structure of $DyB_{12}$. Red arrow shows the expected direction of dynamic charge stripes in the *fcc* $REB_{12}$ structure; **(b)** schematic view on measurements of the magnetoresistance angular dependence; ***n***- normal vector to the lateral surface of the studied crystal, ***H***- magnetic field, ***I***- measuring current and $\varphi$ is the angle between ***n*** and ***H*** vectors. Temperature dependences of **(c)** magnetization $4\pi M(T, H_0$=100 Oe) and resistivity $\rho(T)$, and of **(d)** specific heat $C(T, H_0)$ for ***H***||[112]. AF and P denote the antiferromagnetic and paramagnetic phases, $T_N$ and $T_M$ are the temperatures of magnetic phase transitions.

**III.2. *H-T* diagrams for the principal *H* directions**. The suppression of the AF state by external magnetic field has been studied by magnetization and magnetoresistance measurements for the principal magnetic field directions ***H***||[001], ***H***||[110], ***H***||[111] and ***H***||[112] in crystals of $DyB_{12}$. The temperature dependences of heat capacity $C(T, H_0)$ were also measured for ***H***||[112] orientation (Fig. 1d). Figs. 2 and 3 show the field dependences of magnetic susceptibility

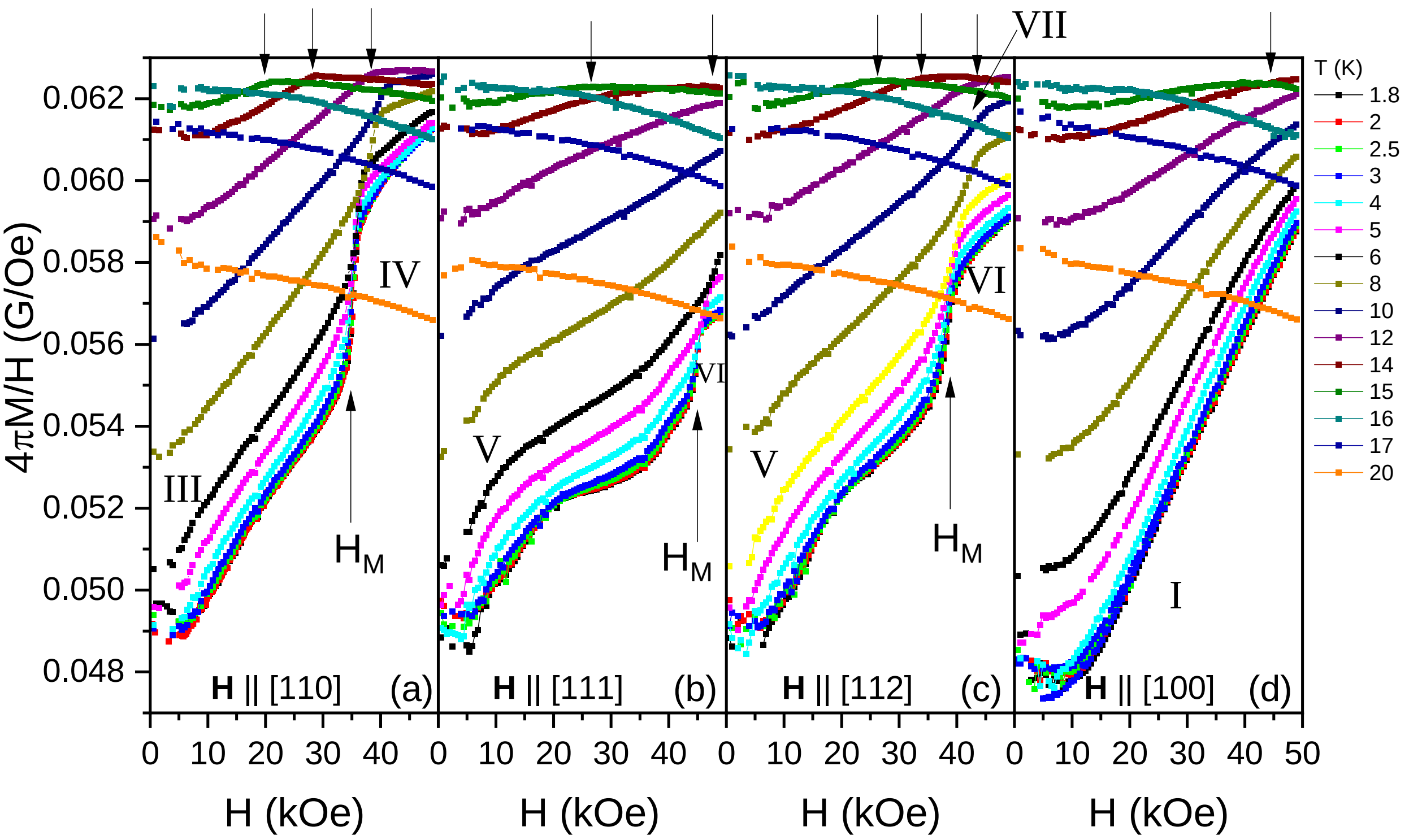

**Fig. 2.** Magnetic field dependences of magnetic susceptibility $4\pi M/H(H, T_0)$ of $DyB_{12}$ at fixed temperature $T_0 \leq 20$ K for principal directions ***H***||[110] (a), ***H***||[111] (b), ***H***||[112] (c) and ***H***||[100] (d). Arrows mark the magnetic phase transitions in the AF state and between AF and P phases, Roman numerals denote the magnetic phases (see Fig. 4).

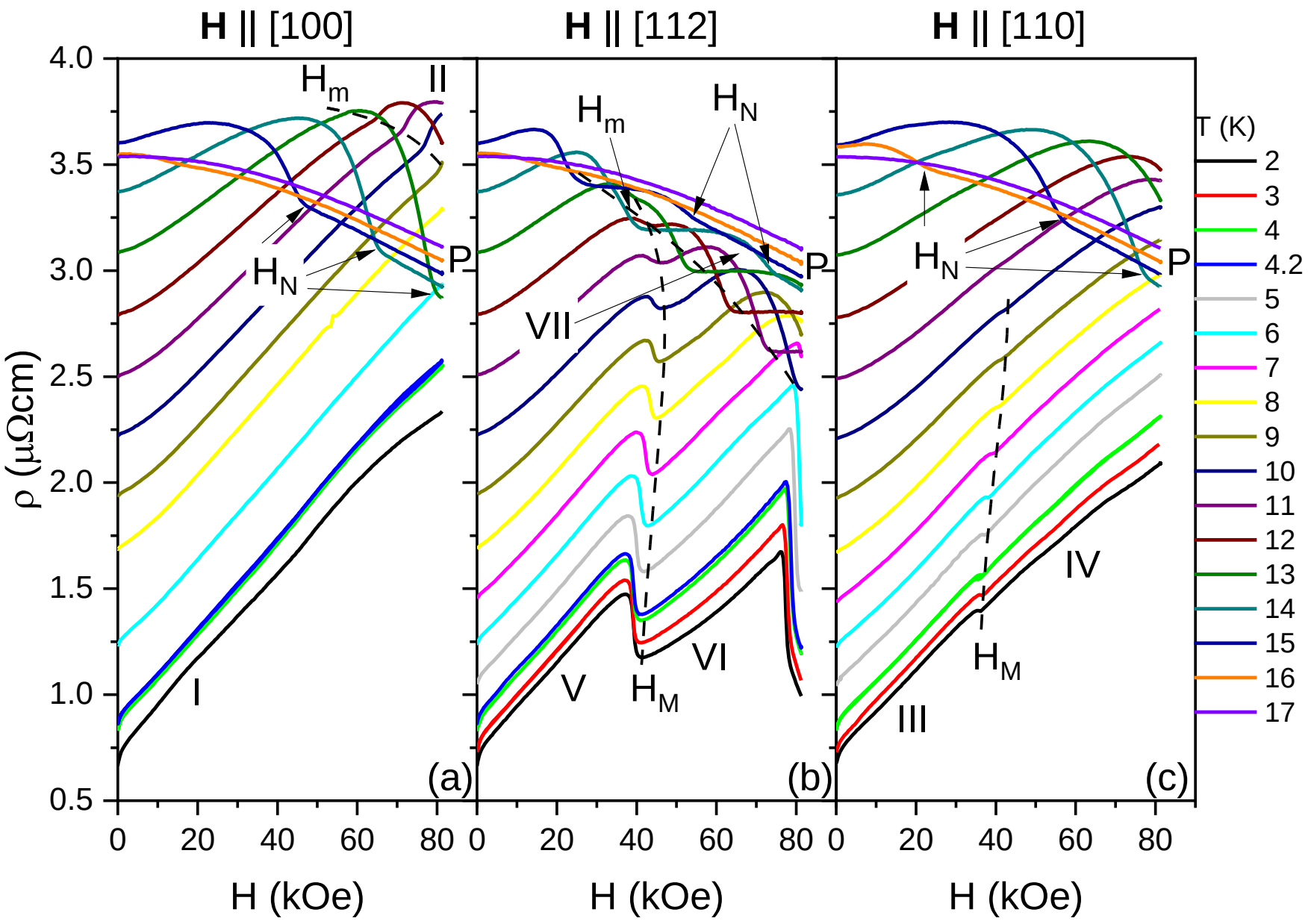

**Fig. 3.** Magnetic field dependences of resistivity $\rho(H,T_0)$ of $DyB_{12}$ at fixed temperature $T_0 \leq$ 17 K for principal field directions ***H***||[100] (a), ***H***||[112] (b) and ***H***||[110] (c). Arrows mark the magnetic phase transitions in the AF state and between AF and P phases, Roman numerals denote the magnetic phases (see Fig. 4).

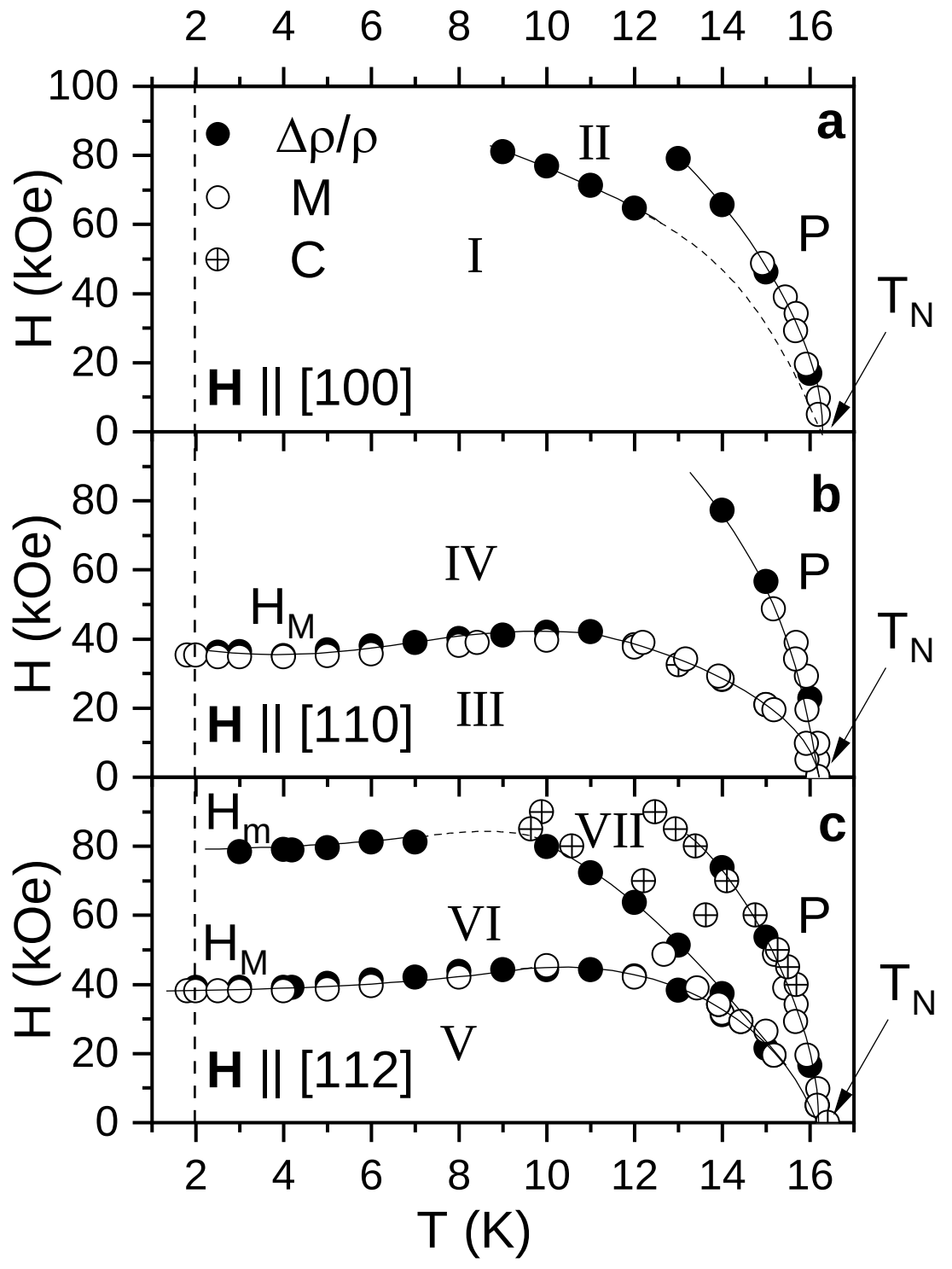

**Fig. 4.** Magnetic *H-T* phase diagrams of $DyB_{12}$ along three principal directions of the external magnetic field (a) ***H***|| [100], (b) ***H***|| [110] and (c) ***H***|| [112]. Roman numerals denote different magnetically ordered phases, P- paramagnetic state. Vertical dashed line marks $T_0$ = 2 K, which corresponds to the results of angular measurements (Figs. 5-6) and to the *H-φ* diagram in the (110) plane (Fig. 7).

$4\pi M$/H($H$, $T_0$) and resistivity $\rho(H,T_0)$, respectively. Arrows mark the magnetic phase transitions in the AF state and between AF and P phases, Roman numerals denote the magnetic phases in Figs. 2-3. Field induced magnetic anisotropy (<7% at $T_0$ = 2 K) is shown for principal directions ***H***||[001], ***H***||[110], ***H***||[111] and ***H***||[112] (see also Fig. S1 in the Supplemental Material [22]). We measured also the magnetization temperature dependences $4\pi M/H(H_0, T)$ at fixed magnetic field to clarify the location of AF-P phase boundary (see Fig. S2 in the Supplemental Material [22]). The obtained *H-T* magnetic phase diagrams are shown in Fig. 4. It will be discussed below that in case of $DyB_{12}$ the principal direction ***H***||[111] matches in the (110) plane the radial phase boundary (see Fig. 8a) along which a strong critical scattering appears. This does not allow the MR and magnetization data analysis of the critical fields developed in the study. As a result, we present in Fig. 4c the *H-T* magnetic phase diagram for ***H***||[112] instead of ***H***||[111].

**III.3. *H-φ* magnetic diagram at $T_0$ = 2 K.** Despite of a moderate anisotropy (~16% at $T$ = 14 K) of the AF-P phase boundary (Fig. 4), the location and the number of magnetic phases inside the AF state is found to be quite different for various ***H*** directions. Testifying the origin of these changes, the angular and field dependences of magnetic susceptibility $4\pi M$/H($\varphi_0$, $H_0$, $T_0$) and magnetoresistance (MR) $\Delta\rho/\rho(\varphi_0, H_0, T_0) = (\rho(H) - \rho(H{=}0))/\rho(H{=}0)$ have been studied at fixed temperature $T_0$ = 2 K (marked with a vertical dashed line in Fig. 4) for various ***H*** directions in the (110) plane. Fig. 5 and Fig. 6 show the sets of the angular resolved magnetic susceptibility and MR dependences, respectively. In more detail, left panels in Figs. 5 and Fig. 6 demonstrate the angular curves at various external magnetic field, the right panels present field dependences for different ***H*** directions, when the orientation of vector ***H*** fixed in the plane (110) (angle $\varphi_0$ is counted from [001], see sketch in Fig. 1b). It can be discerned from Figs. 5-6, that (*i*) magnetic and charge transport anisotropy appears in the range $H$ > 20 kOe, and it increases strongly above 40 kOe, (*ii*) sharp changes on the angular dependences of $4\pi M/H(\varphi_0, H_0)$ and $\Delta\rho/\rho(\varphi_0, H_0)$ allow to detect phase transitions both near <111> axes and in a wide vicinity of <100> directions (radial phase boundaries in the (110) plane), and (*iii*) step-like anomalies of two types are observed in the intervals 35-45 kOe and 75-80 kOe (circular phase boundaries).

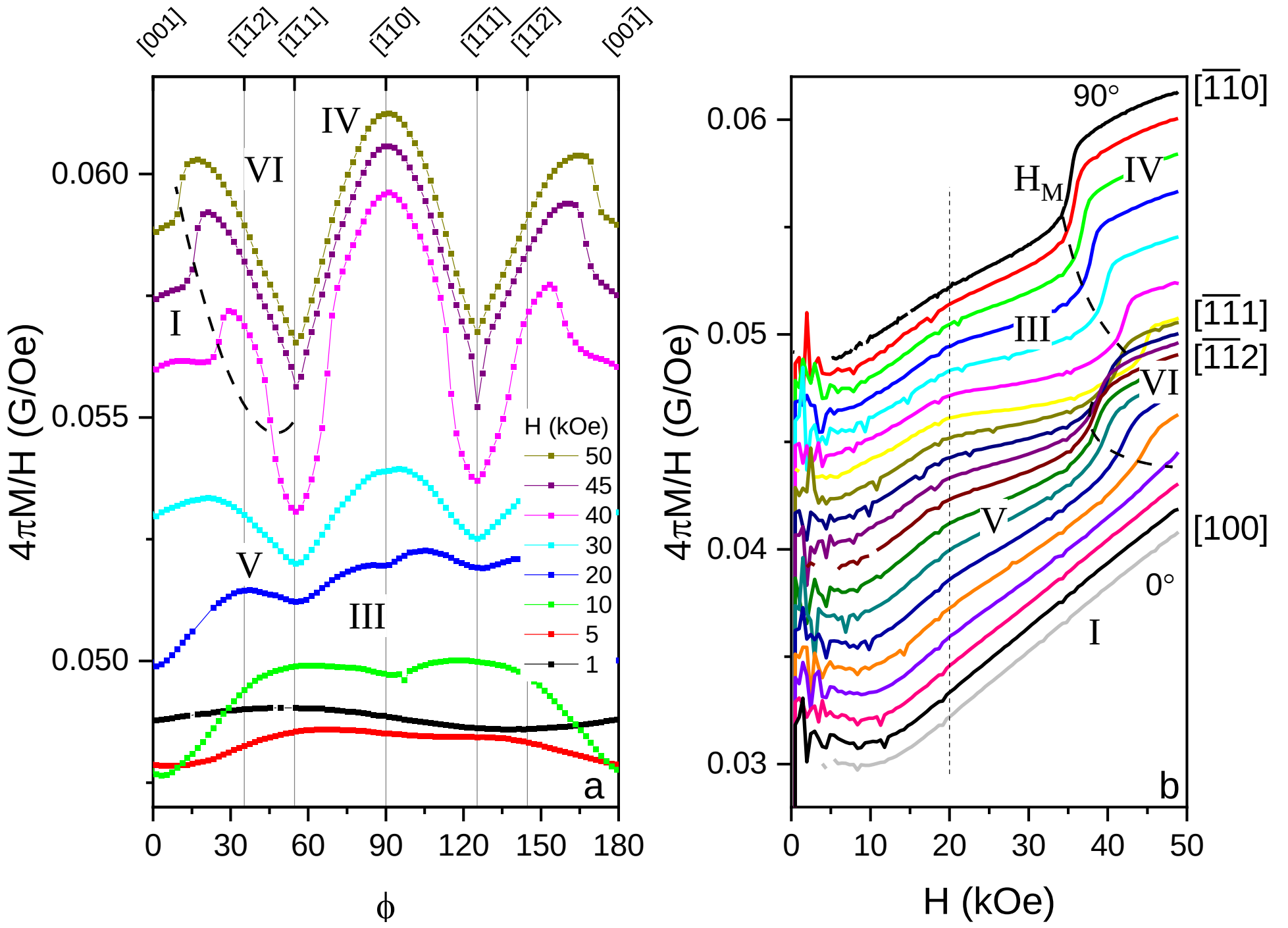

**Fig. 5.** (a) Angular and (b) magnetic field dependences of magnetic susceptibility for the various magnetic fields $H \leq 50$ kOe at $T_0 = 2$ K. The rotation was performed around the axis [110] (angle $\angle\varphi = \boldsymbol{H}$^$\boldsymbol{n}$ counted from [001]). Vertical lines in (a) show positions when the magnetic field $\boldsymbol{H}$ in the (110) plane is aligned with the principal axes in the *fcc* lattice. Curves on (b) are recorded at fixed angles $\angle\varphi$ = [001] ^$\boldsymbol{H}$ in the range 0-90º between [100] and [110] axes (step $\Delta\varphi$=7˚) and shifted for convenience. Roman numerals denote the magnetic phases (see Fig. 4).

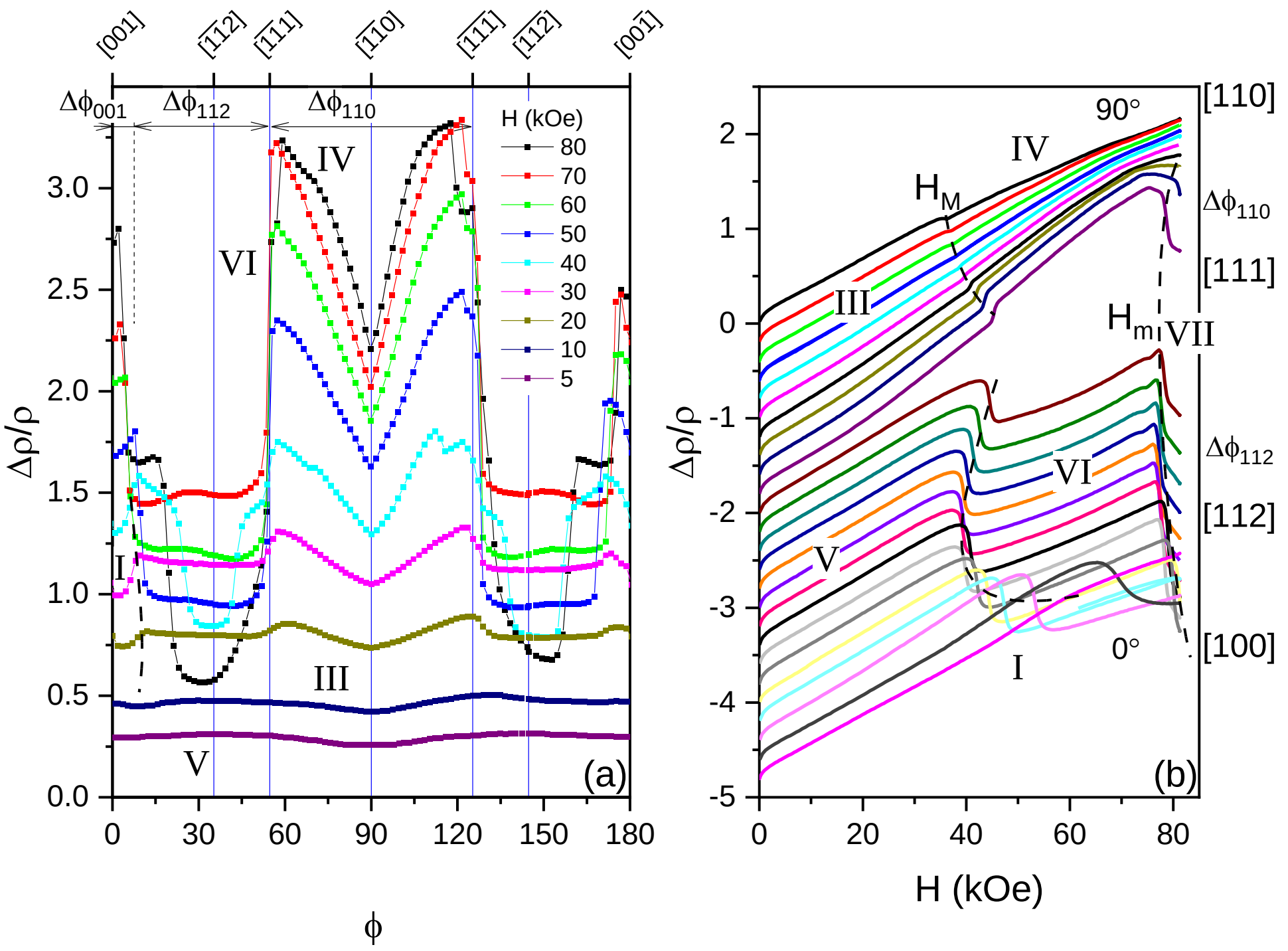

**Fig. 6.** (a) Angular and (b) magnetic field dependences of magnetoresistance for the various magnetic fields $H \leq 80$ kOe at $T_0 = 2$ K. The rotation was performed around the direct current axis [110] (see sketch in Fig. 1b). Vertical lines in (a) show positions when the magnetic field $\boldsymbol{H}$ in the (110) plane is aligned with the principal axes. Curves on (b) are recorded at fixed angles $\angle\varphi$ = [001]^$\boldsymbol{H}$ in the range 0-90º between [100] and [110] axes (step $\Delta\varphi$=3.6˚) and shifted for convenience. Roman numerals denote the magnetic phases (see Fig. 4).

It should be pointed out that the anomalies on the angular magnetic susceptibility $4\pi M/H(\varphi_0, H_0)$ and magnetoresistance $\Delta\rho/\rho(\varphi_0, H_0)$ curves appear at the same directions of $\boldsymbol{H}$ (see, Figs. 5a and Fig. 6a) demonstrating synchronous changes of these two characteristics, and these singularities should be attributed to orientation phase transitions. The MR in the AF state is positive and has an about linear magnetic field dependence below 30 kOe (Fig. 6b). Above 40 kOe a quadratic MR($H$) component appears along with the linear one, and $\Delta\rho/\rho(\varphi_0, H)$ becomes strongly anisotropic distinguishing three main angular segments $\Delta\varphi_{001}$, $\Delta\varphi_{110}$ and $\Delta\varphi_{112}$ (see Figs. 6a-6b) located around three principal crystallographic directions and separated one from another by abrupt radial and circular borders (see the analysis of linear and quadratic $\Delta\rho/\rho(H,T_0)$ components in the section III.4 below). Firstly, the field induced MR anisotropy appears mainly due to the large magnitude anomaly at $H_M$ ~30-40 kOe and then, the changes increase near the second singularity at $H_m$ ~ 80 kOe (Fig. 6b). An overall view of the results of $\Delta\rho/\rho(H,\varphi)$ measurements at $T_0$ = 2 K presented in Figs. 6a-6b is displayed in Fig. 7 in cylindrical coordinates where the MR is plotted along the vertical axis and is additionally provided with a color scale. The projections of these MR (Figs. 6-7) and magnetic susceptibility (Fig. 5) data sets onto the (110) plane (see Figs. 8a and 8b) allow us to refine the $H$-$\varphi$ AF phase diagram at $T_0$ = 2 K. Roman numerals in Fig. 8 show different magnetic I-VII phases in the AF state. Note, that the "butterfly pattern" of the MR anisotropy detected here for $DyB_{12}$ for the first time (Fig. 8a), is similar to that observed for $ErB_{12}$ [12], and it looks quite different from the "Maltese cross" type picture of $\Delta\rho/\rho(H,\varphi,T_0$=2K) obtained both in $HoB_{12}$ [10] and $TmB_{12}$ [13]. The main differences based on the position of the radial phase boundaries on $H$-$\varphi$ diagrams, which are arranged along <111> directions for $DyB_{12}$ ($ErB_{12}$) on the contrary to the case of $HoB_{12}$ ($TmB_{12}$), where the radial borders are located along <112>, limiting the width of the sectors $\Delta\varphi_{110}$ and $\Delta\varphi_{001}$ (Fig. 8a).

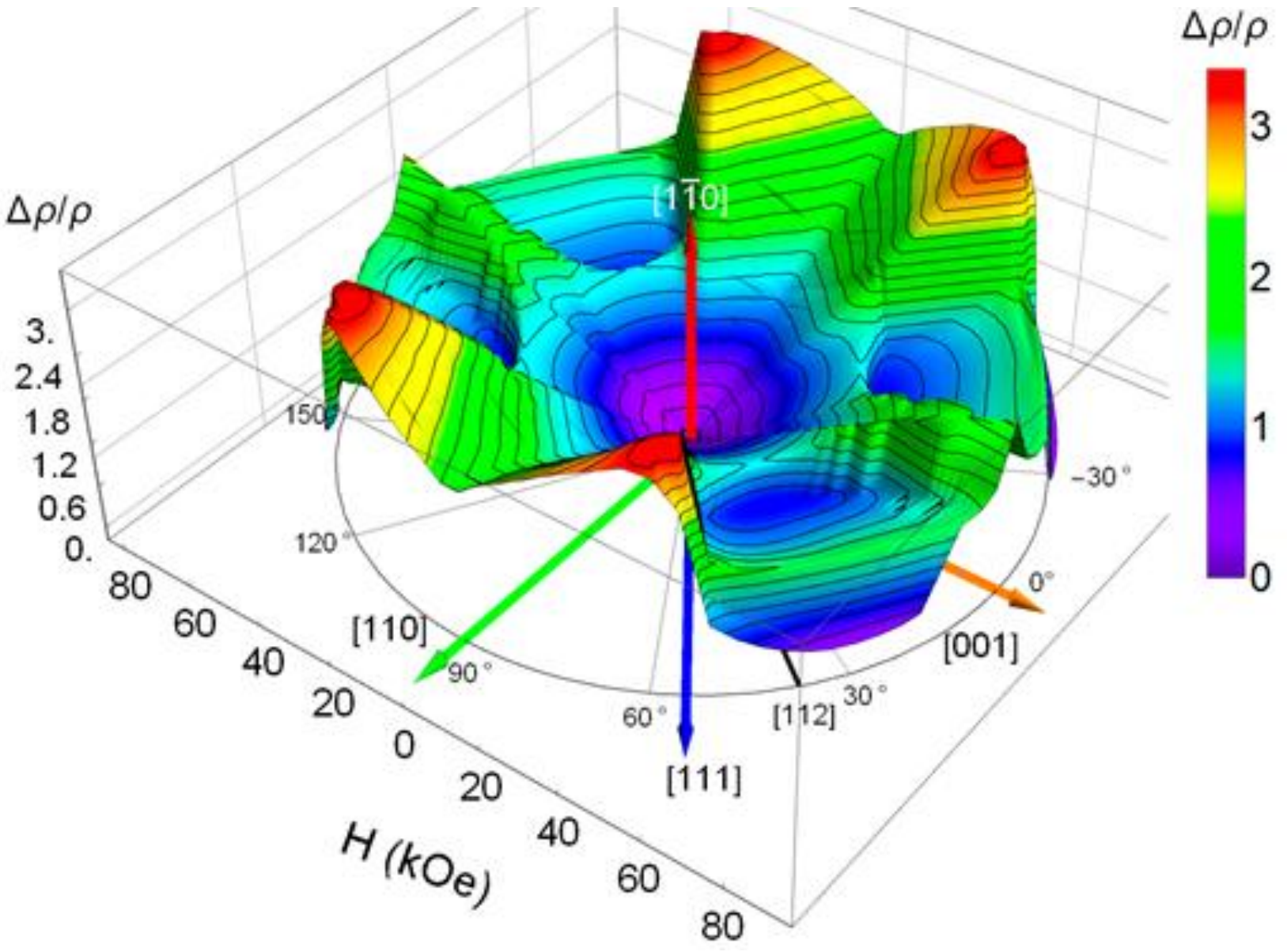

**Fig. 7.** Transverse magnetoresistance $\Delta\rho/\rho = f(H,\varphi)$ in the AF phase of $DyB_{12}$ in cylindrical coordinates for magnetic field rotation in the (110) plane at $T_0$ = 2 K (measuring current $\boldsymbol{I}$||[110]).

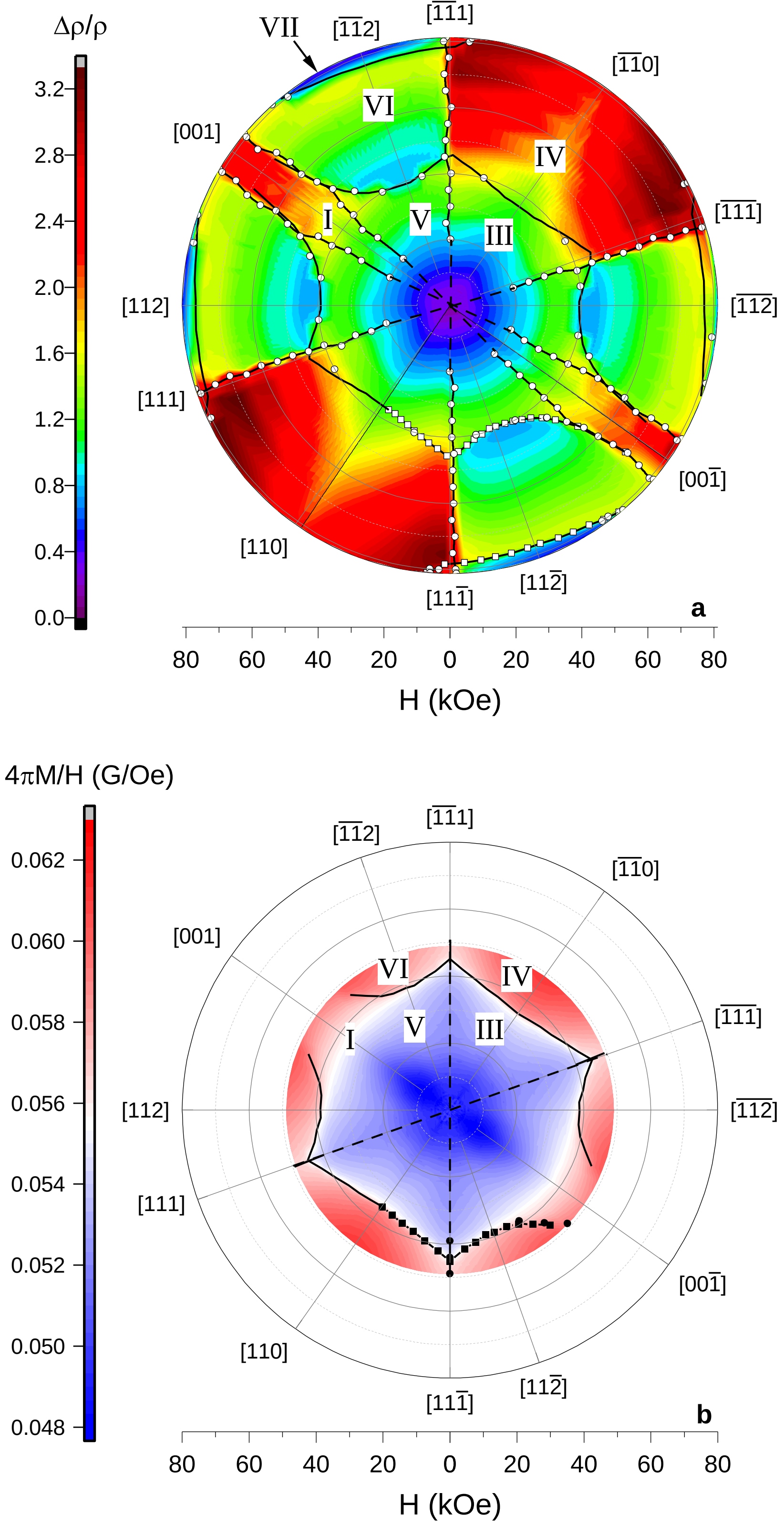

**Fig. 8.** (a) Magnetoresistance of $DyB_{12}$ in polar ($H$, $\varphi$) coordinates in projection onto the (110) plane recorded at $T_0 = 2$ K ($H \leq 80$ kOe, see text). (b) Polar plots of magnetic susceptibility in the range up to 50 kOe for $T_0 = 2$ K**.** Color shows the magnitude of MR and $M/H$ parameters. Phase boundaries are shown as dots and lines, Roman numerals present different magnetic phases in the AF state, which are identical to these shown in Fig. 4.

### III.4. Magnetoresistance components in the AF-phases of $DyB_{12}$.

The magnetic field dependences $\rho(H,\varphi_0,T_0)$ (Fig. 3) and $\Delta\rho/\rho(H,\varphi_0,T_0)$ (Fig. 6b) were used below to separate the positive and negative contributions to magnetoresistance in the AF phases of $DyB_{12}$. Data analysis is very similar to that one developed for $HoB_{12}$, $ErB_{12}$ and $TmB_{12}$ in [23]. As can be seen from Figs. 3 and 6b, the linear positive magnetoresistance dominates in $DyB_{12}$ in ranges $0 < H < H_M$ (denoted below as interval $\Delta H_L$) and $H_M < H < H_m$ (interval $\Delta H_M$, see, for example, Fig. S3 in the Supplemental Material [22]). Slight deviations from the linear MR behavior are observed both in the small-field hysteresis zone $H < 4$ kOe and in the vicinity of transitions at $H_M$, $H_m$ (see derivatives $d\rho/dH(H)$ in Fig. S3 in the Supplemental Material [22]) and small quadratic positive and negative MR components are distinguished additionally in the $\Delta H_{L,M}$ intervals. Besides, a large quadratic negative MR is detected as the main contribution in the range $H_m < H < H_N$ (denoted as $\Delta H_H$, $H_N$ is the Neel field) (see derivatives $d\rho/dH(H)$ in Fig. S4 in the Supplemental Material [22]). Accordingly, in the MR data analysis we used approximations:

$$\frac{\Delta\rho}{\rho} = \begin{cases} A_L H - \frac{B_L}{2} H^2, & when\ H \in \Delta H_L \\ A_M(H - H_M) - \frac{B_M}{2}(H - H_M)^2, & when\ H \in \Delta H_M \\ A_H(H - H_m) - \frac{B_H}{2}(H - H_m)^2, & when\ H \in \Delta H_H \end{cases} \quad (1),$$

where $H_M$ and $H_m$ are the borders between $\Delta H_L$, $\Delta H_M$, and $\Delta H_M$, $\Delta H_H$ intervals, correspondingly, and $A_{L,M,H}$ and $B_{L,M,H}$ are the coefficients in (1) (see Figs. S3 and S4 in the Supplemental Material [22]). Here we took into account that changes in MR should be measured from the low-field boundary of the corresponding region $\Delta H_i$.

Temperature dependences of the $A_{L,M,H}(T/T_N)$ and $B_{L,M,H}(T/T_N)$ coefficients obtained from approximation (1) for $DyB_{12}$ in different intervals $\Delta H_L$, $\Delta H_M$ and $\Delta H_H$ and for three principal directions $\boldsymbol{H}\|[001]$, [110] and [112] are presented in Fig. 9. Positive $A_i$ coefficients are associated usually with the charge carriers scattering on spin-density waves (SDW), whose antinodes are considered as a 5*d* component of the complex antiferromagnetic structure, composed of magnetic moments of the *4f* and *5d* electron states [9-11, 24–25]. As can be seen from Figs. 9a, the SDW associated charge carriers scattering in the AF state decreases gradually when the sample is heated, with $A_i(T)$ zeroing at $T_N$ (at AF-P transition). Besides, over the entire temperature range, the inequality $A_L > A_M$ is valid at least for $\boldsymbol{H}\|[112]$, i.e. in this field direction the amplitude of charge carriers scattering on SDW decreases noticeably at $H_M$ = 35-45 kOe indicating SDW weakening above $H_M$ (Fig. 9a). Additionally we observe a $A_M(T)$ anomaly at $T^* \sim 5$ K, and below $T^*$ the difference between $A_L$ and $A_M$ for the field direction $\boldsymbol{H}\|[112]$ increases strongly. Simultaneously, the positive quadratic component which is characterized by the coefficient $B_M(T)$ in Eq. (1) increases significantly in the range $T < T^* \sim 5$ K (Fig. 9c), and magnetic susceptibility curves $4\pi M/H(T, H_0)$ at $H_0$~40-50 kOe ~ $H_M$ demonstrate a shallow maximum at $T^*$ ~ 5-6 K for field directions $\boldsymbol{H}\|[112]$ and $\boldsymbol{H}\|[111]$ (see Figs. S2c-S2d in [22]). We propose the emergence of a second phase transition at $T^*$ ~ 5-6 K for these field directions, which occurs in vicinity of $H_M \sim$ 40 kOe (see *H-T* diagram in Fig. 4c) and is accompanied by SDW transformation that results in the reduction of SDW scattering. Note that the positive quadratic magnetoresistance $\frac{\Delta\rho}{\rho}(H) \sim B_{L,M}H^2$ (see Eq. (1)) detected in the AF state of $DyB_{12}$ (Fig. 9c) cannot be described in terms of usual Lorentz force induced magnetoresistance, because the estimated $B_{L,M}$ coefficients are too large and lead to a drift mobility of charge carriers $\mu_D$ ~1000-3000 cm$^2$/V s in the relation $\Delta\rho/\rho \approx \mu_D^2 H^2$. On the contrary, a similar liquid helium temperature value $\mu_D$ ~1600 cm$^2$/V s was observed previously for the non-magnetic reference compound $LuB_{12}$ [26], but both the magnetic scattering on Dy-ions and the location of $DyB_{12}$ in vicinity of spinodal boundary [27] have to be considered as two factors that should inhibit significantly the carriers' mobility. As a result, we propose that the negative coefficients $B_{L,M}$ (Fig. 9c) describe a magnetic field-induced slight variation of the main linear positive MR term inside the AF-phases of $DyB_{12}$. A moderate decrease of the linear coefficients $A_{L,M}$ of the MR component in direction ***H***||[112], which is observed in the vicinity of phase transition at $H_M$ (Fig. 9a), may be attributed not only to changes in the structure of SDW at the border between intervals $\Delta H_L$ and $\Delta H_M$, but also to the emergence of spin-polarized electron states in the conduction band. Such a ferromagnetic component in the magnetic structure was observed directly both by neutron diffraction in $HoB_{12}$ [28] and in MR measurements of $RB_{12}$ (R - Ho, Er, Tm) where a significant linear negative ferromagnetic contribution to magnetoresistance has been detected [23, 13-14]. Note, that the drop between $A_L$ and $A_M$, observed at $H_M$ in the series of $RB_{12}$, is the largest for $TmB_{12}$ where the linear term in MR becomes negative, i.e. $A_M < 0$ [23].

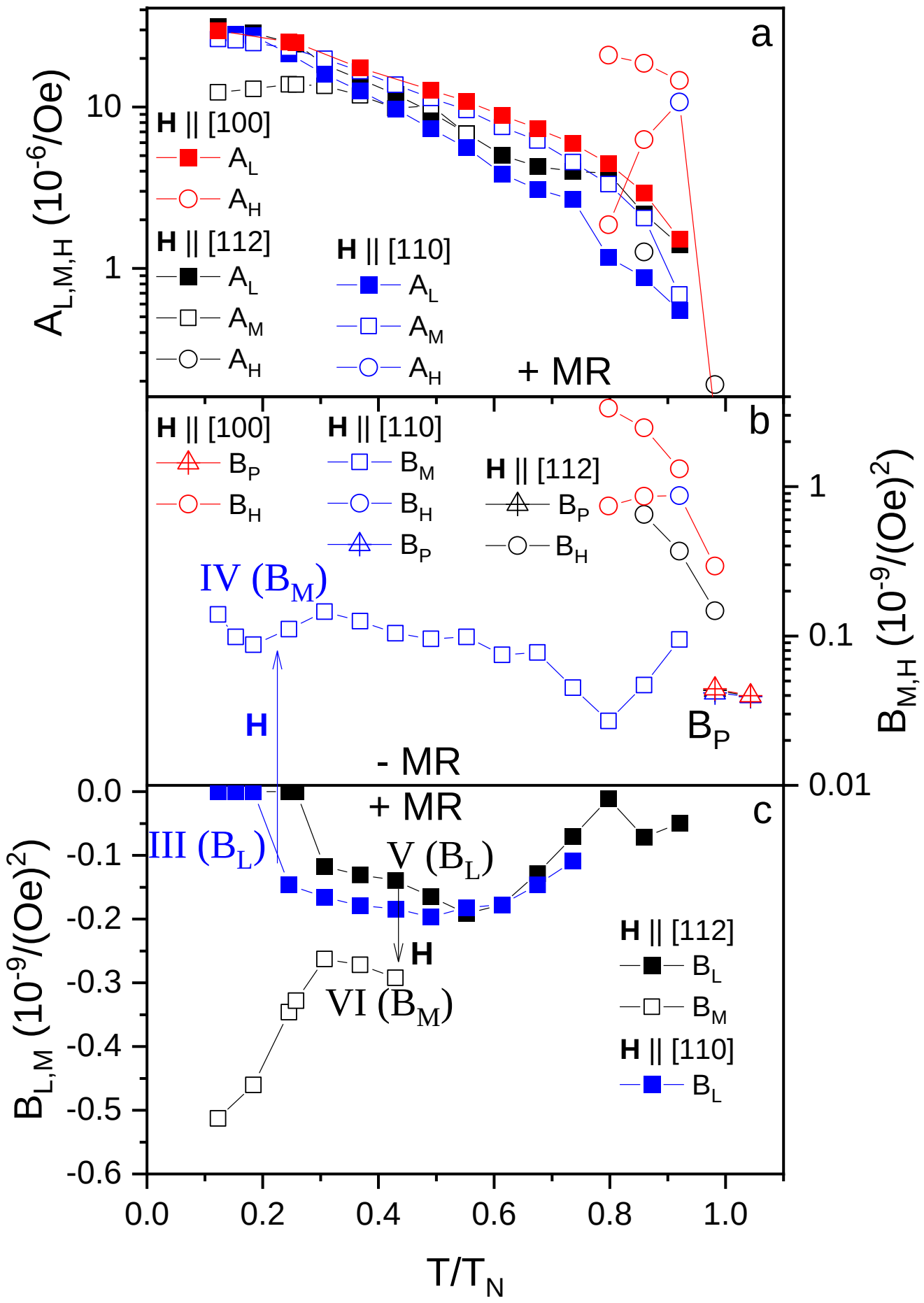

**Fig. 9.** Temperature dependences of parameters $A_{L,M,H}$ (a) and $B_{L,M,H}$ (b-c) for $DyB_{12}$ obtained from a linear approximation of derivatives $\frac{\partial}{\partial H}\left(\frac{\Delta\rho}{\rho}\right)$ in the framework of Eq. (1) for different intervals $\Delta H_L$, $\Delta H_M$ and $\Delta H_H$ of the external magnetic field in AF-state.

It is worth noting that on the contrary to the drastic decrease of MR at $H_M$ observed for ***H***||[112], in the case of ***H***||[110] the amplitude of SDW and correspondingly the linear MR coefficient are enhanced by external magnetic field. As a result, the inequality $A_L < A_M$ is valid for ***H***||[110], and simultaneously a transformation of the additional quadratic MR term from positive to negative values can be detected during the magnetic orientation transition at $H_M$ (Figs. 9b-9c). Only moderate $B_M$ values describe the negative quadratic component, and these are close to the negative coefficient $B_P$ observed here in the paramagnetic state near $T_N$ (Fig. 9b). In the upper interval of magnetic fields $\Delta H_H$ the negative quadratic term $-\frac{\Delta\rho}{\rho}(H)\sim B_H \cdot H^2$ is much higher (Fig. 9b) demonstrating a strong increase of local *4f-5d* spin fluctuations in the vicinity of AF-P phase

transition in $DyB_{12}$ [9-14, 18]. Note, that in $DyB_{12}$ the coefficients $B_P(DyB_{12})$·~ 0.4*$10^{-10}$ $Oe^{-2}$ in the paramagnetic state and $B_H(DyB_{12})$·~ 0.3*$10^{-8}$ $Oe^{-2}$ in the AF-state just below the AF-P transition (Fig. 9b) turn out to be 3-4 times smaller than similar values $B_P(RB_{12})$~1-1.5*$10^{-10}$ $Oe^{-2}$ and $B_H(RB_{12})$·~ 1-2*$10^{-8}$ $Oe^{-2}$ observed in [23] for the magnetic dodecaborides with Ho, Er and Tm ions. Taking into account that $B_P$ and $B_H$ are the characteristics of non-magnetic spin-polarons (heavy fermions) produced by local *4f-5d* spin fluctuations [9-14, 18], a significant increase of these parameters along the $RB_{12}$ series may be attributed to enhancement of the quantum mechanical instability of the 4f-shell, which is growing from $DyB_{12}$ towards the mixed valence narrow-gap semiconductor $YbB_{12}$ [3, 27, 29]. On the contrary, the low temperature values of the linear coefficients $A_{L,M,H}$ (Fig. 9a) are about 1.5 times higher in $DyB_{12}$ than in the series $HoB_{12}$-$TmB_{12}$ [23]. We suppose that the moderate increase of $A_{L,M,H}$ may be explained by the larger magnetic moment of the $\Gamma_{83}$ quartet ground state of Dy-ion in comparison with the other antiferromagnets $RB_{12}$ (R - Ho, Er, Tm) together with weaker 4*f*-5*d* spin fluctuations in $DyB_{12}$ [4]. It is worth noting also a few singularities at $T$ ~ 0.8$T_N$ on $A_L(T)$ and $B_{L,M}(T)$ curves (Fig. 9), but these anomalies are small enough and could not analyzed carefully. Previously, in $TmB_{12}$ and $ErB_{12}$, a ferromagnetic component of magnetic structure was observed below $T_C = 2.5$ K and $T_C \sim 5.4$ K (both values are about $0.8T_N$) [23], correspondingly.

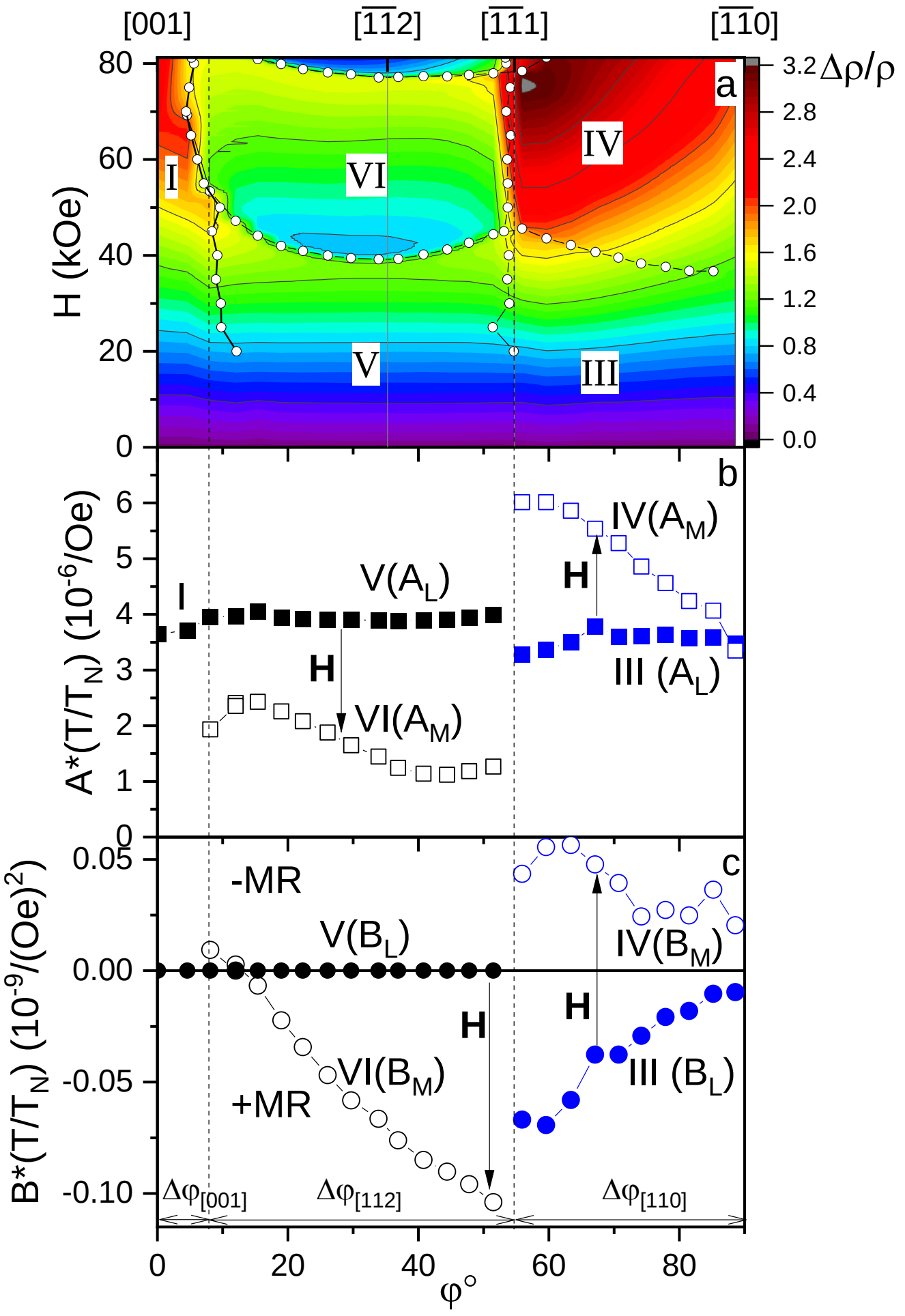

**Fig. 10.** Magnetoresistance coefficients $A_{L,M,H}$ (a) and $B_{L,M,H}$ (b) obtained from a linear approximation of derivatives $\frac{\partial}{\partial H}\left(\frac{\Delta\rho}{\rho}\right)$ in the framework of Eq. (1) for different ***H*** values in intervals $\Delta H_L$, $\Delta H_M$, and $\Delta H_H$ as a dependence on angles $\varphi = (\boldsymbol{n},\boldsymbol{H})$ between the normal to the sample ***n***||[100] and magnetic field ***H***. Rotation at $T$ = 2 K was performed around the electric current direction ***I***||[110] (values $A$ and $B$ are normalized by $T_N$). Roman numerals indicate different AF phases in the (110) plane separated by radial phase boundaries (see vertical dotted lines). Arrows show the change of $A$ and $B$ parameters with increasing magnetic field $H$.

To shed light on the nature of the MR contributions in various magnetic phases of $DyB_{12}$, measurements of magnetic field dependences $\frac{\Delta\rho}{\rho}(H)$ at fixed temperature $T_0 \approx 2$ K were carried out, smoothly changing step-by-step the direction of the external magnetic field $\boldsymbol{H}$ in the (110) plane (rotation of $DyB_{12}$ around the [110] axis with angle $\varphi = \angle(\boldsymbol{H},\boldsymbol{n})$, $\boldsymbol{n}$ - normal vector to the sample lateral surface). Coefficients $A_{L,M}$ and $B_{L,M}$ deduced from the MR analysis in framework of Eq. (1) and normalized on $(T/T_N)$ are shown in Figs. 10b-10c and compared with the $H$-$\varphi$ phase diagram in the plane (110) presented in Fig. 10a. As can be seen from the comparison of panels (a-c), the behavior of $A_{L,M}$ and $B_{L,M}$ in $DyB_{12}$ is strictly tied to the AF phase changes. Indeed, in the entire range of angles 5–54° (sector $\Delta\varphi_{[112]}$ in Fig. 10) an increase of $\boldsymbol{H}$ (indicated by arrows in Fig. 10) induces a *decrease of* coefficient $A$ near $H_M$ from $A_L$ (phase V) to $A_M$ (phase VI), and, on the contrary, the MR slope *increases* from $A_L$ (phase III) to $A_M$ (phase IV) in the interval 54-90° (sector $\Delta\varphi_{[110]}$), as it was previously detected in Fig. 9 at various temperatures for field directions $\boldsymbol{H}$||[112] and $\boldsymbol{H}$||[110]. The angular dependence and the absolute values $1\text{-}6\cdot10^{-6}$ Oe$^{-1}$ of the $A_L$ coefficient (normalized to $T_N$ to be compared with other dodecaborides of the $RB_{12}$ series) in $DyB_{12}$ turn out to be very similar to the other $RB_{12}$ (R - Ho, Er and Tm) [23]. Taking in mind that the butterfly-type $H$-$\varphi$ diagram at $T_0 = 2$ K in $ErB_{12}$ [12] and $DyB_{12}$ (Fig. 8) are similar, it is interesting to compare in more detail the behavior of $A_{L,M}$ and $B_{L,M}$ coefficients in these two AF metals with a $\Gamma_{83}$ quartet ground state of the Kramers-type rare-earth magnetic ions. In sector $\Delta\varphi_{[001]}$ very close normalized values of $A_L(T/T_N)$~3.7 $10^{-6}$ Oe$^{-1}$ and 4.4 $10^{-6}$ Oe$^{-1}$ are observed in $DyB_{12}$ (Fig. 10b) and $ErB_{12}$ [12], respectively (see Fig. S5 in [22] for comparison), and for $ErB_{12}$ the SDW component of MR enhances up to 7 $10^{-6}$ Oe$^{-1}$ during the I-II transition, when the formation of a ferromagnetic contribution to magnetization occurs in phase II [12,23]. In the case of $DyB_{12}$ with $T_N \sim 16.3$ K the phase transition I-II is not observed in magnetic fields up to 80 kOe, so we have not reached the $A_{M,H}$ and $B_{M,H}$ values in sector $\Delta\varphi_{[001]}$. In the interval 54-90° (sector $\Delta\varphi_{[110]}$) the behavior of $A_M$ (phase IV) in $DyB_{12}$ is very close to that one observed for $ErB_{12}$ with $A_M$ variation in the range $3.5\text{-}6\cdot10^{-6}$ Oe$^{-1}$ (Fig. 10a, see Fig. S5 in [22] for comparison). Also quite equal $A_M(\varphi)$ behavior and close values $3.5\text{-}4\cdot10^{-6}$ Oe$^{-1}$ are observed for $DyB_{12}$ and $ErB_{12}$ in the range 5–54° (sector $\Delta\varphi_{[112]}$ in Fig. 10) that argue in favor of a similar magnetic structure and mechanism of charge carriers scattering in these two antiferromagnets with Kramers RE-ions. On the other hand, a dramatic difference in negative $B_M$ coefficients (40-50 times, see Fig. 10c and Fig. S5 in [22]) between $DyB_{12}$ and $ErB_{12}$ in the $\Delta\varphi_{[110]}$ sector and the emergence of an unusual positive quadratic MR in the magnetic phases III and VI (Fig. 10c) likely may be attributed to weaker 4*f*-5*d* on-site spin fluctuations and emergence of very strong disorder in $DyB_{12}$ located in vicinity of the spinodal boundary in the $RB_{12}$ series [27].

To summarize, the MR and magnetization anisotropy in $DyB_{12}$ is mainly determined by different characteristics in three regions: (*i*) in the region along the [100] directions (transverse to the dynamic charge stripes directed along <110>, see Fig. 1a), including phases I and II (Fig. 4a), (*ii*) in a wide range of angles near the <110> directions (the expected direction of dynamic charge stripes according to [3, 6, 8-14]) restricted by the radial boundaries at $\boldsymbol{H}$||[111] (see phases III and IV in Figs. 4b and 8) and (*iii*) in a wide vicinity of the directions <112> (phases V, VI and VII in Figs. 4c and 8). Such a strong anisotropy of MR, magnetization, and angular magnetic phase diagram in the (110) plane (Fig. 8), which is similar to that observed in $ErB_{12}$ [12], supports the formation of the stripe-induced nanoscale phase separation in $DyB_{12}$. Assuming the applicability of the general mechanism proposed in [10-14], the strong anisotropy of magnetoresistance may be attributed to the redistribution of the electron density from RKKY oscillations of the electron spin density to dynamic charge stripes that leads to renormalization (suppression) of the RKKY indirect exchange interaction between nearest $Dy^{3+}$ ions located along the <110> direction of stripes. At the same time, a significant difference between Maltese cross (in the case of $Ho_xLu_{1-x}B_{12}$ and $Tm_{1-}$ ${}_{x}$Yb$_x$B$_{12}$ [9-11,13-14] and butterfly (for $DyB_{12}$ (Fig. 8) and $ErB_{12}$ [12]) patterns of the $H$-$\varphi$ phase diagrams can be attributed to different single-ion anisotropy in the case of non-Kramers (Ho, Tm) and Kramers (Er, Dy) ions. Taking into account that the ground state quartet ($\Gamma_{83}$) in $ErB_{12}$ and $DyB_{12}$ meets the case of the strongest manifestation of anisotropy (the magnitude of the effect at low temperatures is inversely proportional to the excitation energy between two doublets in the quartet), estimations within the framework of the single-ion model and for Dy-Dy pairs are in progress now and the results will be published elsewhere.

## IV. Conclusion

A detailed study of the rather complicated AF state in $DyB_{12}$ dodecaboride with structural (dynamic cooperative Jahn-Teller effect of the boron sub-lattice) and electron instabilities (dynamic charge stripes), which has been performed by precise angle-dependent magnetoresistance and magnetization measurements, shows that the main sectors of the angular $H$-$\varphi$ phase diagram are arranged along directions perpendicular ($\boldsymbol{H}$ || [001]) and parallel ($\boldsymbol{H}$ || [110]) to dynamic charge stripes. The observed strong anisotropy in highly symmetric *fcc* $DyB_{12}$ with a butterfly-type magnetic $H$-$\varphi$ phase diagram exhibiting numerous AF phases is presumably caused by a mechanism associated with the renormalization (suppression) of the RKKY indirect exchange interaction that arises due to the redistribution of the electron density from RKKY oscillations to dynamic charge stripes.

**Acknowledgements.**

The work was partly performed using the equipment of the Shared Research Centre of Lebedev Physical Institute of RAS. S.G. and K.F. acknowledge the support of the Slovak Research and Development Agency under contract No. APVV−23−0226.

# Supplemental material to the article
# Magnetic Phase Diagrams of Antiferromagnet $DyB_{12}$ with Jahn-Teller Lattice Instability and Electron Phase Separation

A.N. Azarevich[1], A.V. Bogach[1], K.M. Krasikov[1], V.V. Voronov[1],
S.Yu. Gavrilkin[2], A.Yu. Tsvetkov[2], S. Gabani[3], K. Flachbart[3], N.E. Sluchanko[1]

*[1]Prokhorov General Physics Institute, Russian Academy of Sciences, Vavilov str. 38, Moscow 119991, Russia*

*[2]Lebedev Physical Institute, Russian Academy of Sciences, Moscow, Leninsky Prospect 53, 119991 Russia*

*[3]Institute of Experimental Physics of the Slovak Academy of Sciences, Watsonova 47, SK-04001 Košice, Slovakia*

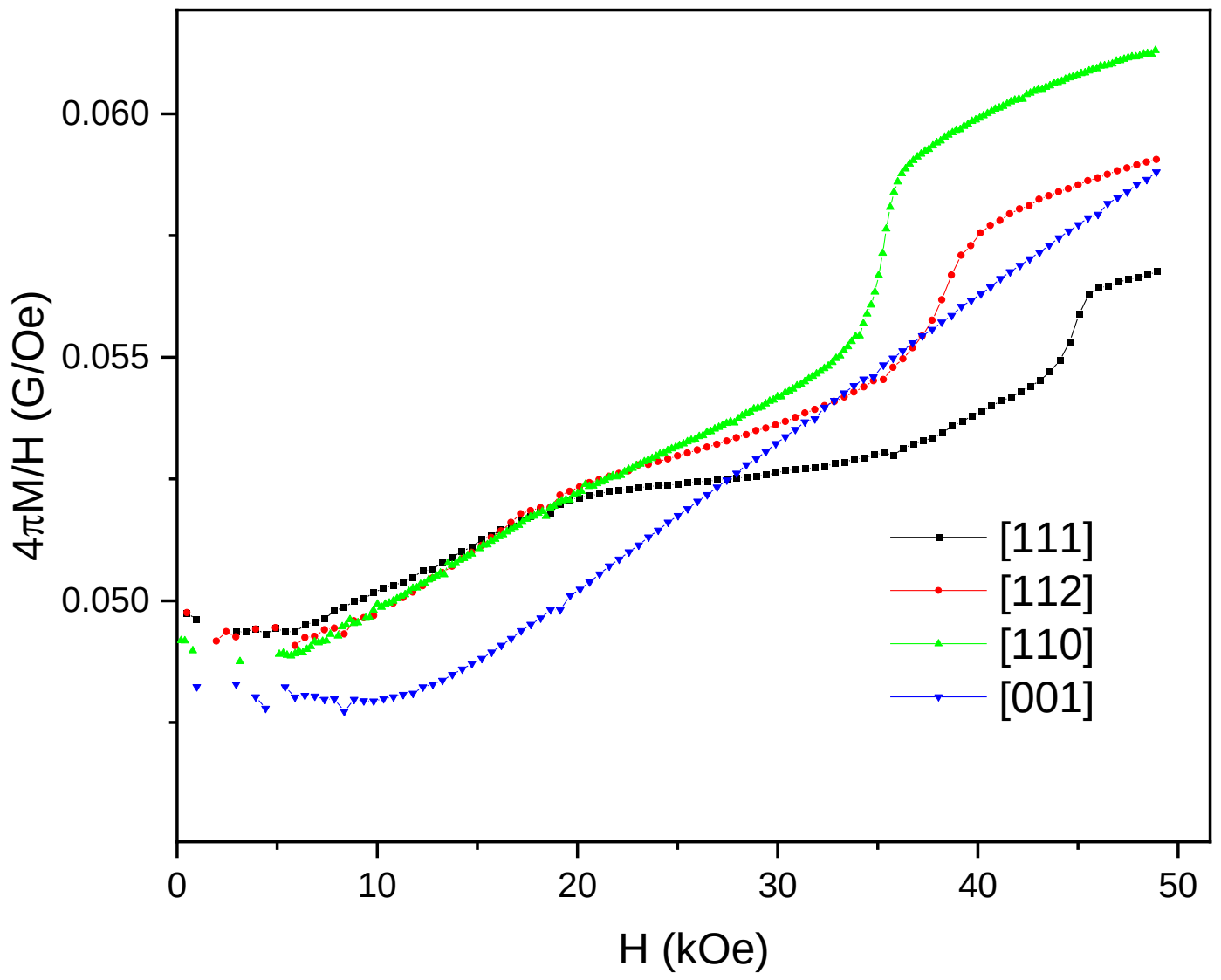

**Fig. S1.** Field induced magnetic anisotropy at $T_0 = 2$ K for principal directions ***H***||[001], ***H***||[110], ***H***||[111] and ***H***||[112] in $DyB_{12}$.

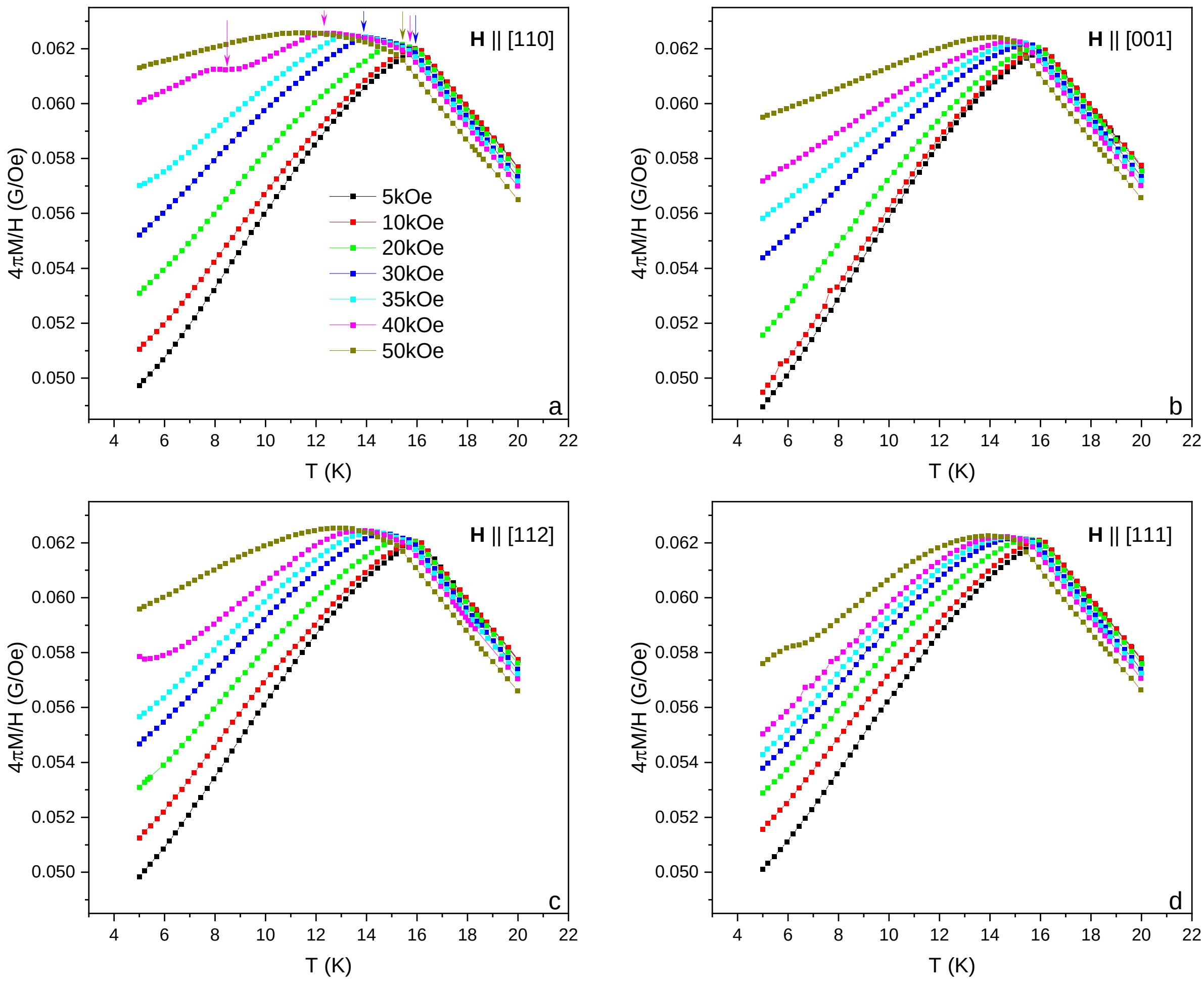

**Fig. S2.** Temperature dependences of magnetic susceptibility $4\pi M/H(H_0, T)$ at fixed magnetic field $H_0 \leq 50$ kOe in directions $\boldsymbol{H}$||[001], $\boldsymbol{H}$||[110] and $\boldsymbol{H}$||[112]. Arrows mark the magnetic phase transitions in the AF state and between AF and P phases.

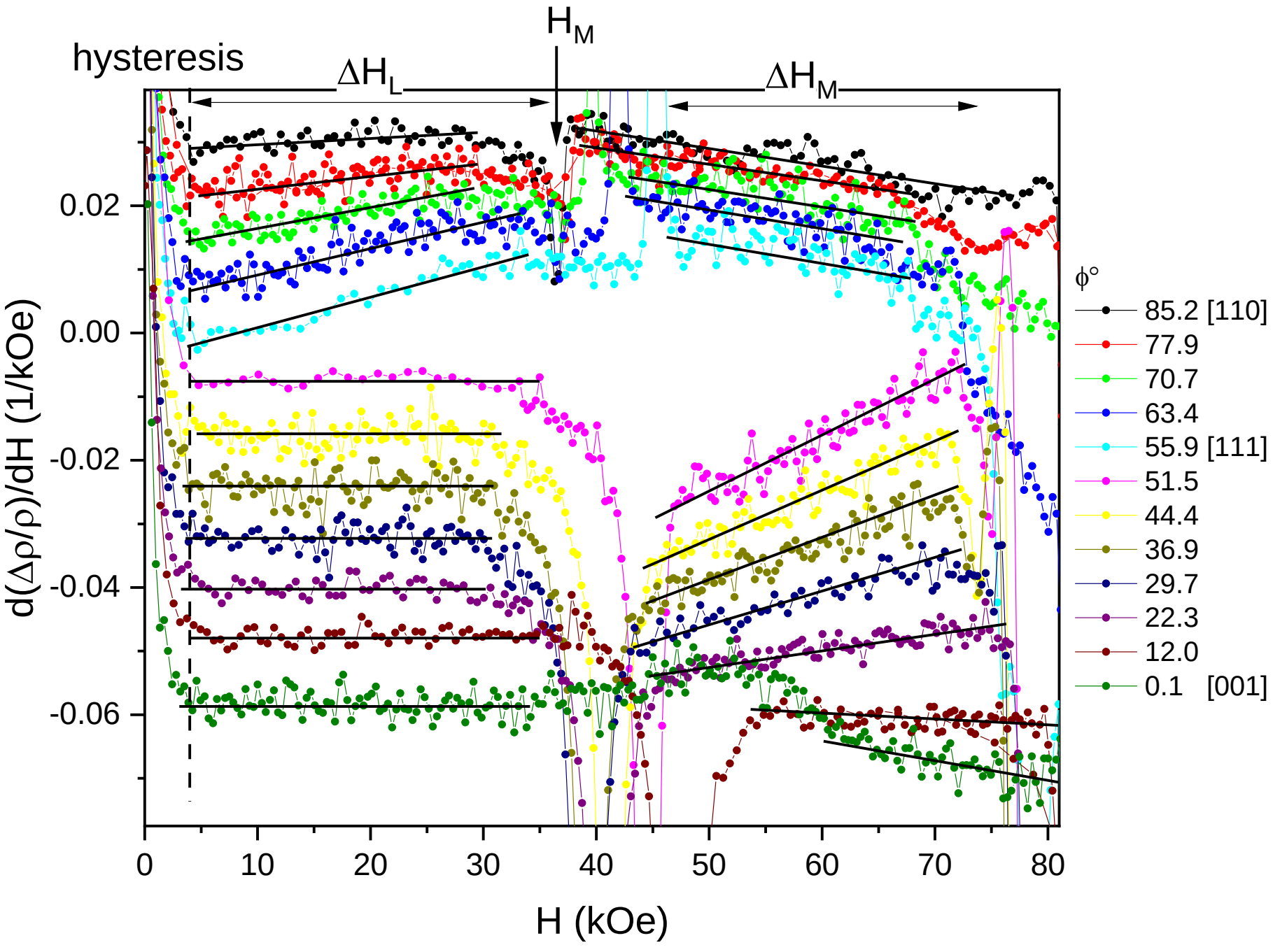

**Fig. S3.** Magnetic field dependences of resistivity derivatives d$\rho$/d$H$ at $T_0$ = 2K for $\boldsymbol{H}$||(110). Marks $\Delta H_L$ and $\Delta H_M$ denote the magnetic field intervals inside different AF phases below fields $H_M$ and $H_m$ (see text in the paper for more detail). Vertical dotted line (left) limits the hysteresis area and solid lines show the linear approximations.

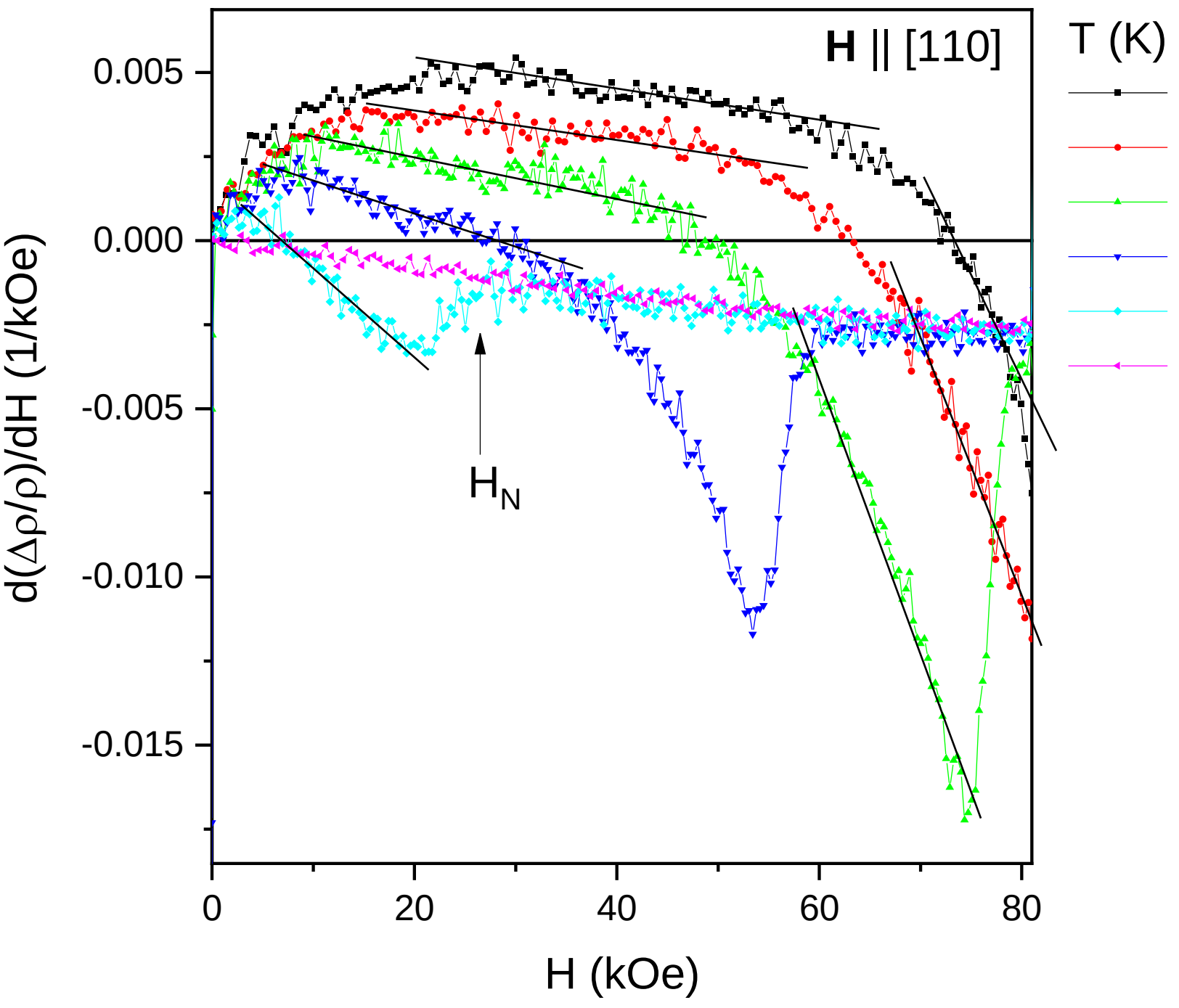

**Fig. S4.** Magnetic field dependences of resistivity derivatives d$\rho$/d$H$ at various temperatures just below $T_N$ for $\boldsymbol{H}$||(110). $\boldsymbol{H_N}$ denotes Néel field.

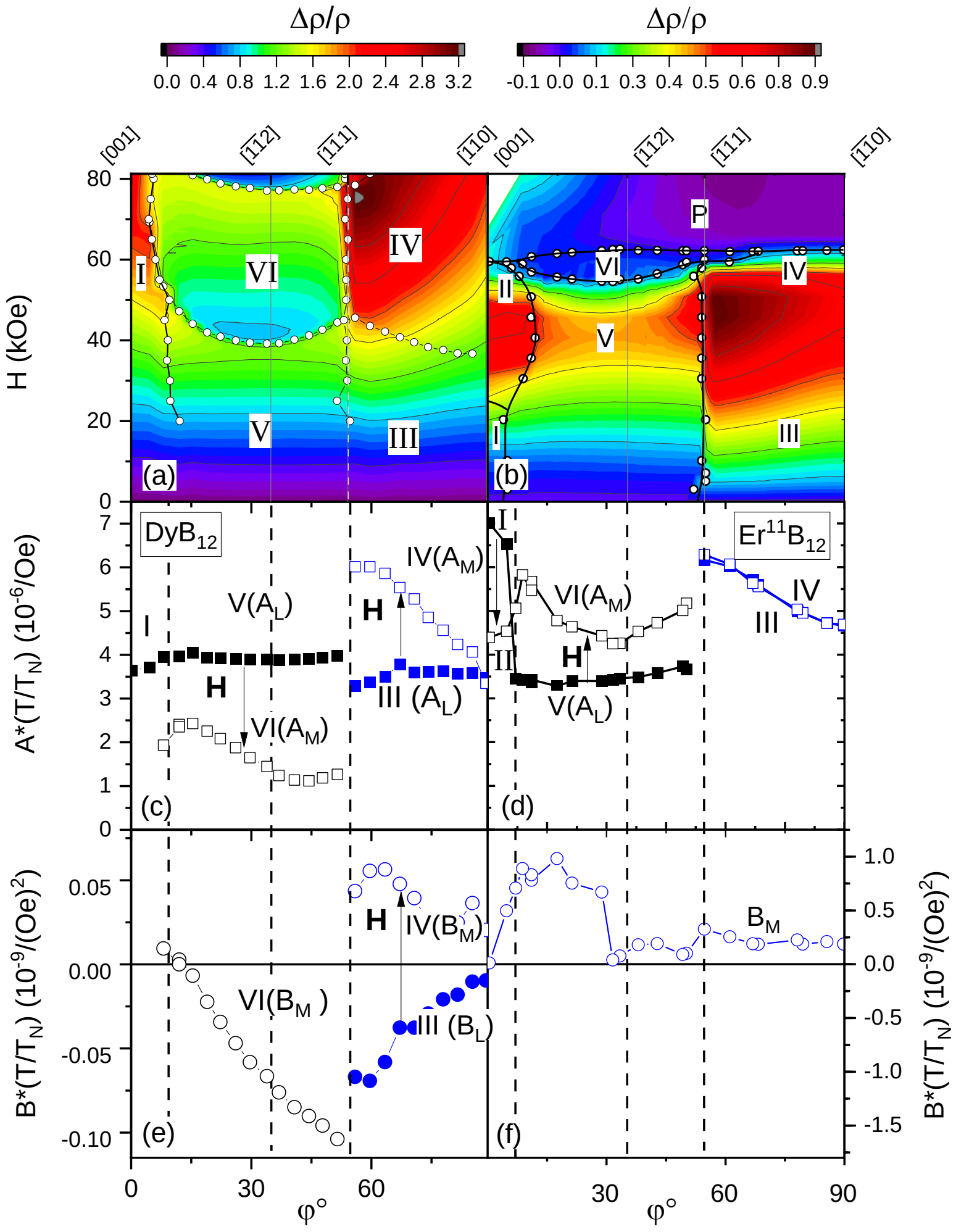

**Fig. S5.** (a) Magnetoresistance $\frac{\Delta\rho}{\rho} = f(H, \varphi)$ of $Er^{11}B_{12}$ for different $\boldsymbol{H}$ values and angles $\varphi$ = ($\boldsymbol{n}$,$\boldsymbol{H}$) between the normal to the sample $\boldsymbol{n}$||[001] and magnetic field $\boldsymbol{H}$. Rotations were performed around current directions $\boldsymbol{I}$||[110]. Parameters $A_{L,M}$ (b) and -$B_H$ (c) for $Er^{11}B_{12}$ obtained from the linear approximation of derivatives $\frac{\partial}{\partial H}\left(\frac{\Delta\rho}{\rho}\right)$ in framework of Eq. (1) for different intervals of the external magnetic field (marked as L, M and H) in AF-phases (values are normalized by $T_N$). Roman numerals indicate different AF phases, P denotes the paramagnetic state. Arrows show the change of the $A$ parameter (from $A_L$ to $A_M$) with increasing magnetic field (b-c).